\documentclass{PoS}

\usepackage{comment}
\usepackage{caption}
\usepackage{subfigure}

\newcommand{\lambdab}{{\overline{\lambda}}}
\newcommand{\chib}{{\overline{\chi}}}

\newcommand{\psib}{{\overline{\psi}}}

\newcommand{\xib}{{\overline{\xi}}}

\def\bec{\begin{center}} 
\def\eec{\end{center}}
\def\beq{\begin{equation}}
\def\eeq{\end{equation}}
\def\bea{\begin{eqnarray}}
\def\eea{\end{eqnarray}}

\title{Higgsing gauge symmetries with reduced staggered fermions}

\ShortTitle{Dynamical Gauge Symmetry Breaking}

\author{Simon Catterall\\
        Department of Physics, Syracuse University\\
        E-mail: \email{smcatterall@physics.syr.edu}}

\author{\speaker{Aarti Veernala}\\
        Department of Physics, Syracuse University\\
        E-mail: \email{aveernal@syr.edu}}

\abstract{We show how a strongly coupled lattice theory consisting of just fermions and 
gauge fields can exhibit a dynamical Higgs mechanism through
the formation of a gauge invariant four fermion condensate.
Furthermore, we argue that the resultant lattice Higgs phase may survive into
the continuum limit.}

\FullConference{31st International Symposium on Lattice Field Theory - LATTICE 2013\\
		July 29 - August 3, 2013\\
		Mainz, Germany}

\begin{document}

\section{Introduction}
The idea that the Higgs mechanism can occur through the formation of fermionic condensates is an
attractive one when constructing many theories of Beyond Standard Model (BSM) physics and finds application in technicolor, composite Higgs models,
tumbling and grand unification schemes
\cite{TC-intro,TC-intro2, Raby:1979my, gut}. 
Lattice realizations of these scenarios thus potentially give a rigorous setting for understanding 
how non-perturbative dynamics in models without elementary scalars can
spontaneously break gauge symmetries and potentially can give us new tools to analyze such theories. 

In a lattice theory, Elitzur's theorem \cite{Elitzur:1975im} guarantees that any
condensate which is not invariant under the gauge symmetry must necessarily have vanishing
expectation value. Instead, the gauge invariant way to understand the operation of the
Higgs mechanism in such theories is that it proceeds via the condensation of a gauge invariant four
fermion operator.  In the models we will consider the gauge interactions factorize
into strong and weak sectors. The Higgsing of the weak gauge symmetry will
be signaled by the appearance of 
a non-zero vacuum expectation for a four fermion operator of the form  
$\phi^\dagger \phi$  where $\phi$ is a composite Higgs
field which is a singlet under the strong interaction but is charged under the
weak group.

In section 2, we start our construction by considering two independent
{\it massless} staggered fermions. Each of these fields can be projected (reduced) so that
one of them is only defined on even parity lattices sites while the other resides on the odd lattice
sites.  
The key observation is that
the kinetic terms for each of these reduced staggered fermions may 
be gauged independently from each other. However, as a consequence, 
single site fermion bilinear terms  will, in general, break
gauge invariance. 
We use this feature to build interesting models
in which the strong interactions force the formation of condensates which Higgs the
weakly coupled symmetries.

\section{Lattice Model and Gauge Symmetry Breaking}

We start with two staggered fields $\chi$ and $\xi$. After restricting them to odd/even sites, we can define new fields $\psi$ and $\lambda$ as :

\begin{eqnarray}
\psib_+(x)=\frac{1}{2}\left(1+\epsilon(x)\right)\chib(x), \; \; \; \; \; \; \; \; \lambdab_-(x)=\frac{1}{2}\left(1-\epsilon(x)\right)\xib(x)\nonumber\\
\lambda_+(x)=\frac{1}{2}\left(1+\epsilon(x)\right)\xi(x), \; \; \; \; \; \; \; \;  \psi_-(x)=\frac{1}{2}\left(1-\epsilon(x)\right)\chi(x)
\end{eqnarray}
where the parity of a lattice site is given by $\epsilon(x)=\left(-1\right)^{\sum_{\mu=1}^4 x_\mu}$.
The fields $\psi$ and $\lambda$ are termed  {\it reduced} staggered fermions since each contains half the
number of degrees of freedom of the usual staggered fermion and corresponds to
two rather than four Dirac fermions in the continuum limit \cite{Smit, Golterman}. The resultant lattice action can then be written as,

\beq
S=\sum_{x,\mu}\eta_\mu(x)\psib_+(x)\left(\psi_-(x+\mu)-\psi_-(x-\mu)\right)+
\sum_{x,\mu}\eta_\mu(x)\lambdab_-(x)\left(\lambda_+(x+\mu)-\lambda_+(x-\mu)\right)
\eeq
where the phase 
$\eta_\mu(x)$ is the usual staggered quark phase given by
\beq  \eta_{\mu}(x) = (-1)^{\sum_{i=1}^{\mu - 1} x_{i}} . \eeq
Since the fields $\lambda$ and $\psi$ in the action are uncoupled we can take them to transform
in different representations
of one or more internal symmetry groups. For example,

\begin{eqnarray}
\psib_+ \to \psib_+ G^\dagger, \; \;  \; \; \; \;\psi_- \to  G\psi_- ,  \; \; \; \; \; \; \lambdab_- \to \lambdab_-H^\dagger , \; \; \; \; \; \; \lambda_+&\to& H\lambda_+
\end{eqnarray}
where $G\in SU(N)$ and $H\in SU(M)$. Notice that these symmetries have a {\it vector} character
in the staggered lattice theory.
These symmetries can be made local by inserting appropriate gauge links
between the $\psi$ and $\lambda$ fields on neighboring sites.
However, notice that unless $G=H$, it is then impossible to write down a single site mass term
that preserves these symmetries.  The usual  staggered mass term
\beq
\psib_+(x)\lambda_+(x)+\lambdab_-(x)\psi_-(x)\eeq  is not invariant.
Throughout the remainder of this paper we will assume
that the interactions factorize into a strongly coupled sector and a weakly coupled sector and that
the $\psi$ and $\lambda$ fields transform differently {\it only under the weak symmetries}.
Indeed, in what follows, we will suppress all indices related to the strong interactions.
In addition, in this talk, we will restrict ourselves to the case where $G=SU(2)$ and $H$ is
trivial. Explicitly we take the $\psi$ field to transform in the fundamental of $SU(2)$ while
the $\lambda$ field is sterile. Both will also transform in the fundamental representation
of a additional strongly coupled $SU(3)$ gauge interaction.

Within this class of models it is possible to write down a gauge invariant lattice four fermion term:
\beq
\sum_{\mu} \phi^\dagger_{+}(x)U_{\mu}(x)\phi_{-}(x),
\eeq 
where the composite Higgs field $\phi(x)$ (an $SU(3)$ singlet) is given by
\begin{eqnarray}
\phi^\dagger_+(x)=\psib_+(x)\lambda_+(x),  \; \; \; \; \; \; 
\phi_-(x)=\lambdab_-(x)\psi_-(x)
\end{eqnarray}
and the weak gauge link $U_\mu$ is needed to render the expression (weak) gauge invariant.
If a condensate of $\phi^\dagger \phi$ develops it will signal the appearance of a lattice phase
in which the weak $SU(2)$ interaction is completely Higgsed.

\section{Details}

We can look for such a condensate using numerical simulations. The gauging of
the lattice kinetic term is given explicitly as
\begin{eqnarray}
S_K=&\sum_{x,\mu}&\psib_+(x)\left(U_\mu(x)V_\mu(x)\psi_-(x+\mu)-U^\dagger_\mu(x-\mu)V^\dagger_\mu(x-\mu)\psi_-(x-\mu)\right)\nonumber \\
&\sum_{x,\mu}&\lambdab_-(x)\left(V_\mu(x)\lambda_+(x+\mu)-V^\dagger_\mu(x-\mu)\lambda_+(x-\mu)\right),
\label{staggered_kinetic}
\end{eqnarray} where, $V_{\mu}(x)$ and $U_{\mu}(x)$ correspond to the strongly coupled SU(3) and the weakly coupled SU(2) gauge groups respectively. In order to see the Higgs phase, we need to add a small perturbation to the above action via an auxiliary field $\phi$. 
\beq
\delta S=g\sum_x \left(\phi(x)\psib_+(x)\lambda_+(x)+\phi^\dagger(x)\lambdab_-(x)\psi_-(x)\right)
\eeq
This field $\phi(x)$ is a local field which is needed to couple the two reduced fermions
together and must transform under
the gauge symmetry as a fundamental of $SU(2)$ in order that the perturbation is
gauge invariant.
To render the path integral well defined after
integration over $\phi(x)$ one must
then also add a suitable action for $\phi(x)$. We choose an additional simple term  $\sum_x \phi^\dagger(x)\phi(x)$.
The effect of these Yukawa  terms is to add a small gauge invariant  four fermion interaction to the action that
favors the conjectured symmetry breaking pattern. 
The lattice action will also contain Wilson plaquette terms for both the $SU(3)$ and
$SU(2)$ gauge fields with corresponding gauge couplings $\beta_S$ and $\beta_W$. 

\section{Results}
It will be important in our
argument to show that
the magnitude of any induced four fermion condensate is insensitive to the value of this 
auxiliary Yukawa coupling $g$ and depends
instead only on the magnitude of the strong gauge coupling. 
\begin{figure}
\begin{center}
\includegraphics[width=0.7\textwidth]{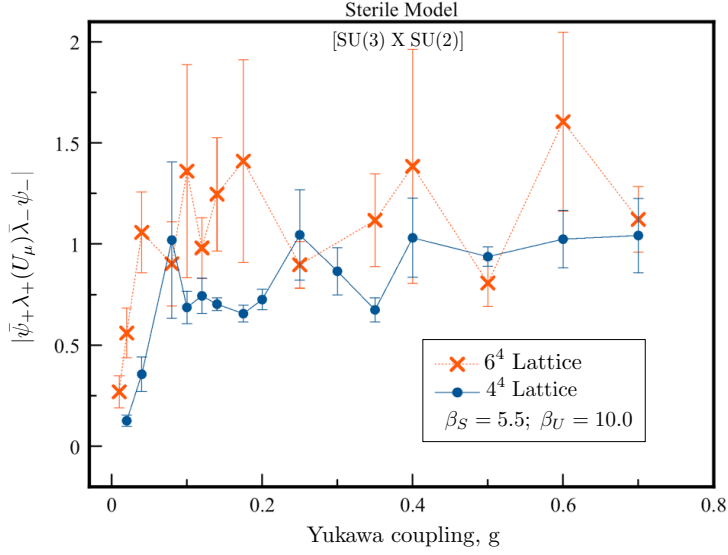}\end{center}
\caption{\label{fig1}Absolute value of the four fermion condensate vs Yukawa coupling $g$}
\end{figure}

\begin{figure}
\begin{center}
\includegraphics[width=0.7\textwidth]{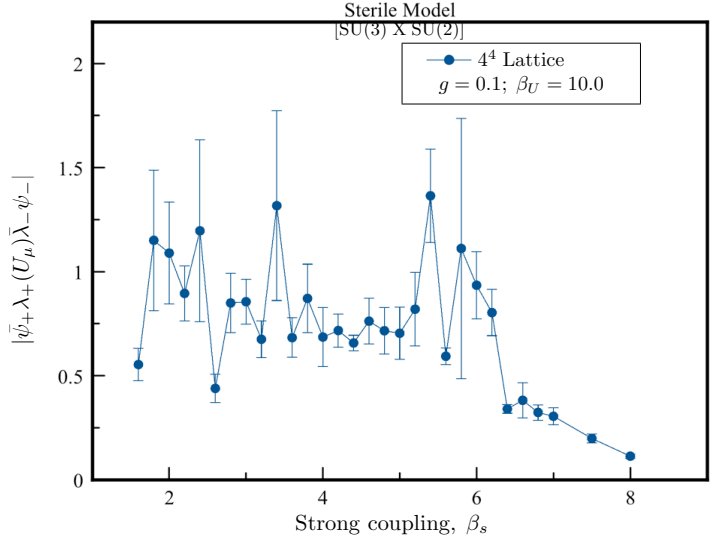}
\end{center}
\caption{\label{fig2} Absolute value of the four fermion condensate vs $\beta_S$}
\end{figure}
Figure~1. shows a plot of the (absolute value of) the four fermion condensate  as a function
of the Yukawa coupling $g$ for fixed $\beta_S=5.5$ (the latter was determined to lie within
the confining phase of the strong interaction for small lattices). We see that indeed a plateau develops and the condensate is insensitive to the value of the Yukawa coupling over a wide
range of $g$.
In contrast figure~2. shows that the measured condensate {\it does} depend on the
strong coupling constant $\beta_S$ (here $g=0.1$ is held fixed). The condensate is
enhanced for small $\beta_S$ and falls towards zero once $\beta_S\ge 6.0$ 
which we observe is large enough to
cause deconfinement in the strong sector. Most
importantly we see no sign of a phase transition as we vary the strong coupling. 
This is evidence that
any Higgs phase of the lattice theory may survive the continuum limit $\beta_S\to \infty$.

The appearance of a Higgs phase can be seen more clearly in figure~3. which shows the
Polyakov line corresponding to the weak gauge field as a function of the strong coupling
$\beta_S$.
For large $\beta_S$ the weak Polyakov line $P_W$ is large
but as $\beta_S$ is lowered it rapidly crosses over to fluctuatate around a much 
smaller but non zero value. Since the weak Polyakov line $P_W=e^{-FT}$ measures the free energy $F$ of
an isolated static quark in the fundamental representation of the weak gauge group a value approaching unity is
associated with a deconfined phase. This is to be expected for a bare weak gauge coupling of
$\beta_W=10.0$ on such a small lattice. Conversely, a confining phase typically
would be associated with a small value of $P_W$. 
What is observed for small $\beta_S$ is intermediate between these two regimes and may be a signal
of a Higgs phase.
Furthermore, the crossover between these two regimes corresponds precisely
to the switching on of the four fermion condensate and the observation of
confinement in the strong sector.
Thus our numerical
results are at least consistent with the dynamical
generation of a non-zero four fermion condensate and the appearance of
a Higgs phase as a result of
strongly coupled dynamics. 

\begin{figure}
\begin{center}\includegraphics[width=0.7\textwidth]{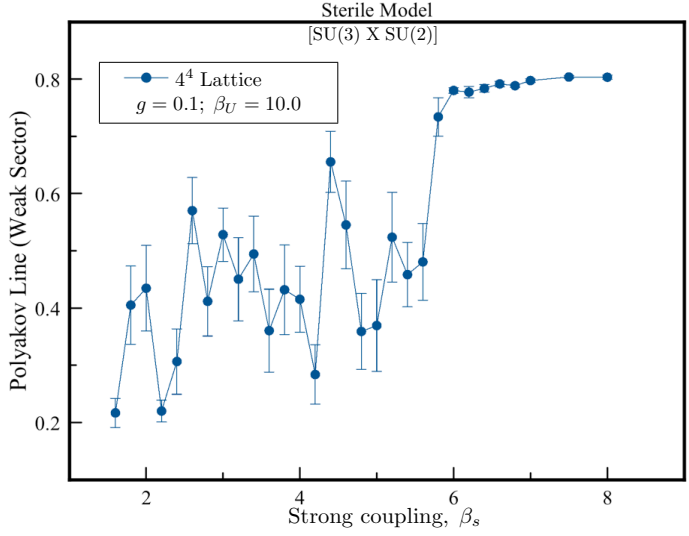}\end{center}
\caption{\label{poly}Weak Polyakov line versus $\beta_S$}
\end{figure}

\section{Conclusions}

We have shown that lattice theories comprising two {\it reduced} staggered fermion fields 
can be constructed in such a way that they can naturally
generate non-perturbative condensates that Higgs weakly coupled gauge symmetries
as a result of strongly coupled dynamics. The key element that allows for symmetry
breaking in these models is the absence of
single site fermion bilinears which are invariant under the weak symmetries.  Since the
vacuum expectation value of such bilinears is always zero the signal for the appearance of
such a Higgs phase is a condensation of a gauge invariant four fermion operator constructed from the
original fermion bilinears. We have illustrated these ideas using a simple model in which
all fields are gauged under a strongly coupled $SU(3)$ interaction while only one of the
fields is gauged under a weak $SU(2)$. 

To understand these results it is instructive to imagine switching off the weak $SU(2)$ interaction
completely. One can then recast the lattice model as a single conventional staggered fermion
comprising the fields $(\psi^1,\lambda)$ together with an additional single reduced
field $\psi^2$ where the explicit indices refer to the global $SU(2)$ symmetry.
For sufficiently strong $SU(3)$ coupling one expects a single site condensate $<\psib^1(x)\lambda(x)>$ to form. Such a condensate will spontaneously break the
global $SU(2)$. Additionally, one expects that the remaining
reduced staggered fermion will also form a condensate of the form
$<\psib^2(x)V_\mu(x) \psi^2(x)>$ where $V_\mu(x)$ is a strong gauge link. Now consider switching on
the weak coupling. This cannot change the nature of the condensates that have
already formed under the strong interaction (indeed this is we mean by weak
coupling) so that one
would conclude that the lattice theory will enter a phase in which the $SU(2)$ symmetry is Higgsed.
The remaining question is whether this lattice Higgs phase survives the continuum limit.

At first glance this seems unlikely since the Vafa Witten theorem prohibits spontaneous
breaking of vector symmetries and the $SU(2)$ has a vector like character when acting on the
original reduced staggered field.
However,  at finite lattice spacing this theorem does not apply since the fermion
measure for this system of reduced staggered fermions is not positive definite. But the
problem would seem to reassert itself for vanishing lattice spacing.  However, in \cite{ourpaper}  we argue
that this symmetry should be interpreted 
as a broken {\it axial} symmetry in the continuum limit. This identification follows
from the usual rules that are used to construct continuum fermions from 
(reduced) staggered fields and implies that such a condensate
would be compatible with the Vafa Witten theorem. It also
violates none of the usual theorems
\cite{Nielson,KS} since the continuum theory is vector-like and trivially free of gauge anomalies.

Notice that this is not the only breaking pattern of symmetry breaking allowed; one could imagine instead that each reduced field forms its own gauge invariant single link condensate and $SU(2)$ is left
unbroken. We see no evidence of this in our numerical work but it is a logical possibility and
indeed what would have been expected for a vector-like continuum theory. The fact that we have both
theoretical arguments and numerical work which support a different scenario is very interesting and
argues that it is important to perform a larger scale study in which the continuum limit can be
examined in more detail.

It should be noted that the question of whether $SU(2)$ breaks after it is weakly gauged
exists even in the continuum; it depends on how $SU(2)$ is
embedded into the global chiral symmetry group where it constitutes the well known vacuum
alignment problem.
We thank
Maarten Golterman and Yigal Shamir for pointing this out \cite{MS}.

In conclusion,
it seems that the lattice models discussed here may serve as useful toy models
for understanding the possibilities for dynamical symmetry breaking in strongly
coupled gauge theories and can be used
to test ideas such as the maximal attractive channel hypothesis 
and tumbling scenarios. It would be interesting
to investigate such models via strong coupling expansions which would avoid possible
sign problems.

\end{document}